\documentclass[%
 reprint,
 amsmath,amssymb,
 aps,
]{revtex4-2}
\pdfoutput=1
\usepackage{dcolumn}
\usepackage{bm}
\usepackage{amsfonts}
\usepackage{amssymb}
\usepackage{amsmath}    
\usepackage{graphicx}   
\usepackage{verbatim}   
\usepackage{color}      
\usepackage[colorlinks=true,linkcolor=blue,citecolor=blue,allcolors=blue]{hyperref}
\usepackage{lineno}
\usepackage{lipsum}
\usepackage{multirow}
\usepackage{afterpage}
\usepackage{flushend}
\usepackage[export]{adjustbox}
\usepackage{float}
\usepackage{footmisc}
\usepackage[english]{babel}
\usepackage[utf8]{inputenc}
\usepackage[section]{placeins}
\makeatletter
\newcommand*{\rom}[1]{\expandafter\@slowromancap\romannumeral #1@}
\makeatother

\begin{document}



\title{Theory of Nonlinear Acoustic Forces Acting on Inhomogeneous Fluids}
\author{Varun Kumar Rajendran,$^1$ \ Sujith Jayakumar,$^1$ \ Mohammed Azharudeen,$^1$ \ and Karthick Subramani$^1$} 
\email{karthick@iiitdm.ac.in.}
\affiliation{$^1$ Department of Mechanical Engineering, Indian Institute of Information Technology, Design and Manufacturing Kancheepuram, Chennai-600127, India.}

\begin{abstract}
Recently, the phenomena of streaming suppression and relocation of inhomogeneous miscible fluids under acoustic fields were explained using the hypothesis on mean Eulerian pressure.  In this letter, we show that this hypothesis is unsound and any assumption on mean Eulerian pressure is needless.  We present a theory of non-linear acoustics for inhomogeneous fluids from the first principles, which explains streaming suppression and acoustic relocation in both miscible and immiscible inhomogeneous fluids inside a microchannel.  This theory predicts the relocation of higher impedance fluids to pressure nodes of the standing wave, which agrees with the recent experiments.
\end{abstract}

\maketitle

\par The acoustic fields imposed on fluids exhibits several interesting nonlinear acoustic phenomena including acoustic radiation force acting on particle or interface, and acoustic streaming \cite{Friend2011Jun}. This subject has long history, beginning with early investigations by Faraday \cite{Michael1831Dec}, Rayleigh \cite{rayleigh}, King \cite{king}, and  Lighthill \cite{lighthill}. Over the last two decades, employing these acoustic forces in microscale flows  has become a rapidly growing research field known as “Acoustofluidics”. It has far ranging applications in biological \cite{Lee2015Mar,Petersson2007Jul,Wiklund2012May, Collins2015Nov, Ahmed2016Mar}, medical \cite{Augustsson2012Sep,Li2015Apr} and chemical sciences \cite{Suslick1999Feb}. Recently, two interesting phenomena were observed when the acoustic fields are imposed on inhomogenous fluid called acoustic relocation \cite{Deshmukh2014Jul} and streaming suppression \cite{Augustsson2016May}. Karlsen et al \cite{Karlsen2016Sep} theoretically explained both the above phenomena using acoustic force density which is divergence of time averaged momentum flux tensor, $\textbf{\emph{f}}_{ac}= -\nabla\cdot\langle\Pi\rangle=-\nabla\cdot\langle p_2I+\rho_0\textbf{\emph{v}}_1\textbf{\emph{v}}_1\rangle$, where $\langle p_2\rangle$ is time averaged second order mean Eulerian pressure and $\nabla\cdot\langle\rho_0\textbf{\emph{v}}_1\textbf{\emph{v}}_1\rangle$ is the time averaged Reynolds stress tensor due to first order acoustic fields. The mean Eulerian pressure is assumed as $\langle p_2\rangle=(1/2)[\kappa_0\langle|p_1|^2\rangle-\rho_0\langle|\textbf{\emph{v}}_1|^2\rangle]$ and this assumption is claimed as the central hypothesis of their theory \cite{Karlsen2018}. Following are our objections to this assumption: First, like any other pressure field, the mean Eulerian pressure has to be derived from Navier - Stokes (N-S) equations. Second, if the assumed $p_2$ is second order Eulerian pressure then the pressure that results from  N-S equations after the substitution of $-\nabla\cdot\langle\Pi\rangle$ lacks clarity. In this letter, we show that the above hypothesis is unsound and unnecessary. 

\par The main goal of this letter is to develop a theory of nonlinear acoustics for inhomogenous fluids that explains the phenomena of acoustic relocation and streaming suppression. Remarkably, our theory predicts that the acoustic relocation/stabilization of inhomogeneous fluids in a microchannel subjected to standing acoustic waves is possible only if there exist an impedance ($Z=\rho c$) gradient which agrees well with the recent experiments. We demonstrate that the amplitude of first order fields are highly dependent on fluid configuration, thus acoustic energy density ($E_{ac}$) varies significantly during the process of relocation. Also, we successfully separate the streaming term and acoustic relocation term from the generalized acoustic body force, which previously claimed not possible \cite{Karlsen2018Jan}. Furthermore, this theory also explains the relocation of immiscible fluids under acoustic fields.

\emph{Physics.} - The hydrodynamics of inhomogeneous fluids considered in this study is governed by mass-continuity, momentum and advection-diffusion equations, \cite{Landau1987Aug} 
\begin{subequations}
\label{Eq 1}
\begin{equation}
\label{Eq 1a}
    \partial_t \rho+\nabla \cdot\left(\rho \textbf{\emph{v}}\right)=0
\end{equation} 
\begin{multline}
\label{Eq 1b}
    \rho[\partial_t \textbf{\emph{v}} + (\textbf{\emph{v}}\cdot\nabla)\textbf{\emph{v}}]=-\nabla p +\eta\nabla^2 \textbf{\emph{v}}
    \\
    +\beta\eta\nabla(\nabla\cdot\textbf{\emph{v}}) + \rho\textbf{\emph{g}}
\end{multline}
\begin{equation}
\label{Eq 1c}
    \partial_t \emph{s}+\textbf{\emph{v}}\cdot\nabla \emph{s} = D\nabla^2\emph{s} 
\end{equation}
where $\rho$ is the density, $\textbf{\emph{v}}$ is the velocity, $p$ is the pressure, $\eta$ is the dynamic viscosity of the fluid, $\xi$ is the bulk viscosity, $\beta=(\xi/\eta)+(1/3)$, ${\emph{s}}$ is the solute concentration and \emph{D} is diffusivity. When the fluid is subjected to acoustic waves, following thermodynamic pressure-density relation in terms of material derivative $(d/dt)=\partial_t+(\textbf{\emph{v}}_1\cdot\nabla)$ is also required \cite{Bergmann2005Jun}, 
\begin{equation}
\label{Eq 1d}
    \frac{d\rho}{d t}=\frac{1}{c^2}\frac{dp}{d t} 
\end{equation}
\end{subequations}
where $c^2=(\partial p/\partial\rho)|_{S}$ and $c$ is adiabatic local speed of sound. 

\par According to perturbation theory, the dependent fields $f$ are decomposed as \cite{Bruus2011Dec}, 
\begin{equation}
\label{Eq 2}
    f=f_0(\textbf{\emph{r}},\tau)+f_1(\textbf{\emph{r}},\tau, t_f)+f_2(\textbf{\emph{r}},\tau)
\end{equation}
where $f_{0}$ is zeroth order (background) fields, $f_1$ is first order time-harmonic acoustic fields $f_1=f_a(\textbf{\emph{r}},\tau)e^{-i\omega t_f}$, actuated at an angular frequency $\omega$ $(\sim1 MHz)$,  and $f_{2}$ are second order fields (in general $f_{2}\ll f_1$). In microscale flows, since the hydrostatic pressure $\rho g H (\sim 1 Pa) \ll p_1 (\sim 10^6 Pa)$, the variation of pressure and velocity fields due to gravity is accounted in the second order effects. Thus, in a quiescent fluid, we take the zeroth order velocity $\textbf{\emph{{v}}}_0=0$, and pressure $p_0 = constant$ ($\nabla p_0 = 0$).  

\par The first order acoustic fields vary in fast time scale $t_f$ ($t_f\sim \frac{1}{\omega} \sim 0.1 \mu s $), whereas the second order hydrodynamic fields varies in slow time scale $\tau$ $(\tau\gg t_f)$. Usually in perturbation theory, $f_0(r,\tau)$ is assumed to be constant for homogeneous fluids. Whereas for inhomogeneous fluids, the variation in background fields ($\rho_0$ and $s_0$) with space as well as slow time scale has to be accounted due to the gravity stratification and second order acoustic effects. As the first order acoustic fields ($f_1$) are sensitive to inhomogeneous configuration ($\rho_0$ and $s_0$) which varies with slow time scale, the amplitude of these acoustic fields are considered to be function of slow time scale ($f_a(\textbf{\emph{r}},\tau)$) unlike homogeneous fluids, where the amplitude of these fields are only function of space ($f_a(\textbf{\emph{r}})$).   
\par The variation of zeroth order fields in fast time scale ($t_f$) can be neglected. Also, since $D\sim \mathcal{O}(10^{-9})$, the diffusion term is negligible in fast time scale, due to which composition of any given fluid particle remains unchanged as it moves. Consequently, the governing equations up to first-order ($t\sim t_f$) reduce to \cite{Bergmann2005Jun} 
\begin{subequations}
\label{Eq 3}
\begin{equation}
\label{Eq 3a}
    \partial_t \rho_1 +\nabla \cdot\left(\rho_0 \textbf{\emph{{v}}}_1\right)=0
\end{equation}
\begin{equation}
\label{Eq 3b}
    \rho_0 \partial_t \textbf{\emph{v}}_1 =-\nabla p_1 + \eta\nabla^2 \textbf{\emph{v}}_1+\beta\eta \nabla(\nabla\cdot\textbf{\emph{v}}_1)
\end{equation}
\begin{equation}
\label{Eq 3c}
    \partial_ts_1+\textbf{\emph{v}}_1\cdot\nabla \emph{s}_0 =0
\end{equation}
\begin{equation}
\label{Eq 3d}
    \partial_t\rho_1+(\textbf{\emph{v}}_1\cdot\nabla)\rho_0=(1/c^2)[\partial_tp_1]
\end{equation}
\end{subequations}

\par The first order fields have a harmonic time dependence, thus time average of these fields are zero. Therefore, first order terms in Eqs.(\ref{Eq 3}) cannot cause any bulk fluid motion. However, the Navier-Stokes equation is non linear and  the above linearized Eqs.(\ref{Eq 3}) are not exact. Hence, proceeding to solve  Eqs.(\ref{Eq 1}) up to second-order ($t\sim\tau$),
\begin{subequations}
\label{Eq 4}
\begin{equation}
\label{Eq 4a}
       \left\langle\partial_t(\rho_0+\rho_2)\right\rangle+\nabla\cdot\langle\rho_1 \textbf{\emph{v}}_1\rangle  + \nabla\cdot\langle\rho_0 \textbf{\emph{v}}_2\rangle = 0
\end{equation}
\begin{multline}
\label{Eq 4b}
 \langle\rho_1 \partial_t \textbf{\emph{v}}_1\rangle + \rho_0\langle(\textbf{\emph{v}}_1 \cdot\nabla)\textbf{\emph{v}}_1\rangle-\langle(\rho_0+\rho_2)\textbf{\emph{g}}\rangle=-\nabla\langle p_2 \rangle \\ + \eta\nabla^2 \langle\textbf{\emph{v}}_2\rangle+
    \beta\eta\nabla(\nabla\cdot\langle\textbf{\emph{v}}_2 \rangle)-\left\langle\rho_0 \partial_t \textbf{\emph{v}}_2\right\rangle
\end{multline}
\begin{equation}
\label{Eq 4c}
    \langle\partial_t (s_0+s_2)\rangle+\langle\textbf{\emph{v}}_2\cdot\nabla s_0\rangle+ \langle\textbf{\emph{v}}_1\cdot\nabla s_1\rangle= D\nabla^2\langle(s_0+s_2)\rangle 
\end{equation}
\begin{multline}
\label{Eq 4d}
    \langle\partial_t(\rho_0+\rho_2)\rangle+\langle(\textbf{\emph{v}}_1\cdot\nabla)\rho_1\rangle+\langle(\textbf{\emph{v}}_2\cdot\nabla)\rho_0\rangle=\\(1/c^2)[\langle\partial_tp_2\rangle+\langle(\textbf{\emph{v}}_1\cdot\nabla)p_1\rangle]
\end{multline}
\end{subequations}
where $\langle...\rangle$ denotes time average over one oscillation period. Since the time average of product of two first order fields is nonzero, they act as a source term for second order fields (slow hydrodynamic time scale). The first two terms on the left side of Eq.(\ref{Eq 4b}) together comprise the divergence of Reynolds stress tensor, $\nabla\cdot\langle\rho_0\textbf{\emph{v}}_1 \textbf{\emph{v}}_1\rangle$.

\par In microscale flows, variation of second order fields with respect to slow time scale is negligible \cite{Friend2011Jun,Bruus2011Dec}. From Eqs.(\ref{Eq 3}), $\rho_1\ll\rho_0$ and $s_1\ll s_0$, thus $\rho_2\ll\rho_0$ and $s_2\ll s_0$. We can also neglect $\langle\textbf{\emph{v}}_1\cdot\nabla s_1\rangle$,$\langle\rho_1\nabla\cdot\textbf{\emph{v}}_1\rangle$ since the first order fields in both these terms are out of phase. Using the above arguments, combining Eqs.(\ref{Eq 4a} and \ref{Eq 4d}),
\begin{equation}
\label{Eq 5}
    \langle\rho_0\nabla\cdot\textbf{\emph{v}}_2\rangle=(1/c^2)\langle\textbf{\emph{v}}_1\cdot\nabla p_1\rangle
\end{equation}

\par Substituting the above relation in Eq.(\ref{Eq 4b}) and analysing the order of magnitude, $\mathcal{O}(\beta\eta\nabla\frac{1}{\rho_0 c_0^2}(\textbf{\emph{v}}_1\cdot\nabla)p_1)\ll\mathcal{O}(\nabla\cdot\langle\rho_0\textbf{\emph{v}}_1\textbf{\emph{v}}_1\rangle)$. As this term, $(1/c^2)\langle\textbf{\emph{v}}_1\cdot\nabla p_1\rangle$ does not contribute in momentum equation, it can be neglected. Hence, Eq.(\ref{Eq 5}) becomes divergence free and second-order flow is incompressible i.e., $\nabla\cdot\langle\textbf{\emph{v}}_2\rangle=0$ (Nyborg's approximation-\cite{Nyborg2005Jun,Baasch2020Jan}). Accounting all the above arguments, the governing equations reduce to
\begin{subequations}
\label{Eq 6}
\begin{equation}
\label{Eq 6a}
    \nabla\cdot\langle\textbf{\emph{v}}_2\rangle=0
\end{equation}
\begin{equation}
\label{Eq 6b}
 -\nabla\cdot\langle\rho_0\textbf{\emph{v}}_1 \textbf{\emph{v}}_1\rangle+\langle\rho_{0}\textbf{\emph{g}}\rangle-\nabla\langle p_2 \rangle  + \eta\nabla^2 \langle\textbf{\emph{v}}_2\rangle=0
\end{equation}
\begin{equation}
\label{Eq 6c}
    \langle\partial_t \emph{s}_0\rangle+\langle\textbf{\emph{v}}_2\cdot\nabla \emph{s}_0\rangle = D\nabla^2\langle\emph{s}_0\rangle
\end{equation}
\end{subequations}

\par The above Eqs.(\ref{Eq 6}) govern the dynamics of inhomogeneous fluids in microscale acoustofluidics and can also be derived by another approach, see Supplemental Material . From Eq.(\ref{Eq 6b}), it is evident that the second-order slow hydrodynamic flows are created due to the divergence of Reynolds stress tensor consisting of product of first order fast acoustic fields. Thus, we introduce the body force due to acoustic fields as
\begin{equation}
\label{Eq 7}
    \textbf{\emph{f}}_{ac}=-\nabla\cdot\langle\rho_0\textbf{\emph{v}}_1 \textbf{\emph{v}}_1\rangle
\end{equation}

\par
\begin{figure}[h!]
  \center
    \includegraphics[width=0.85\linewidth]{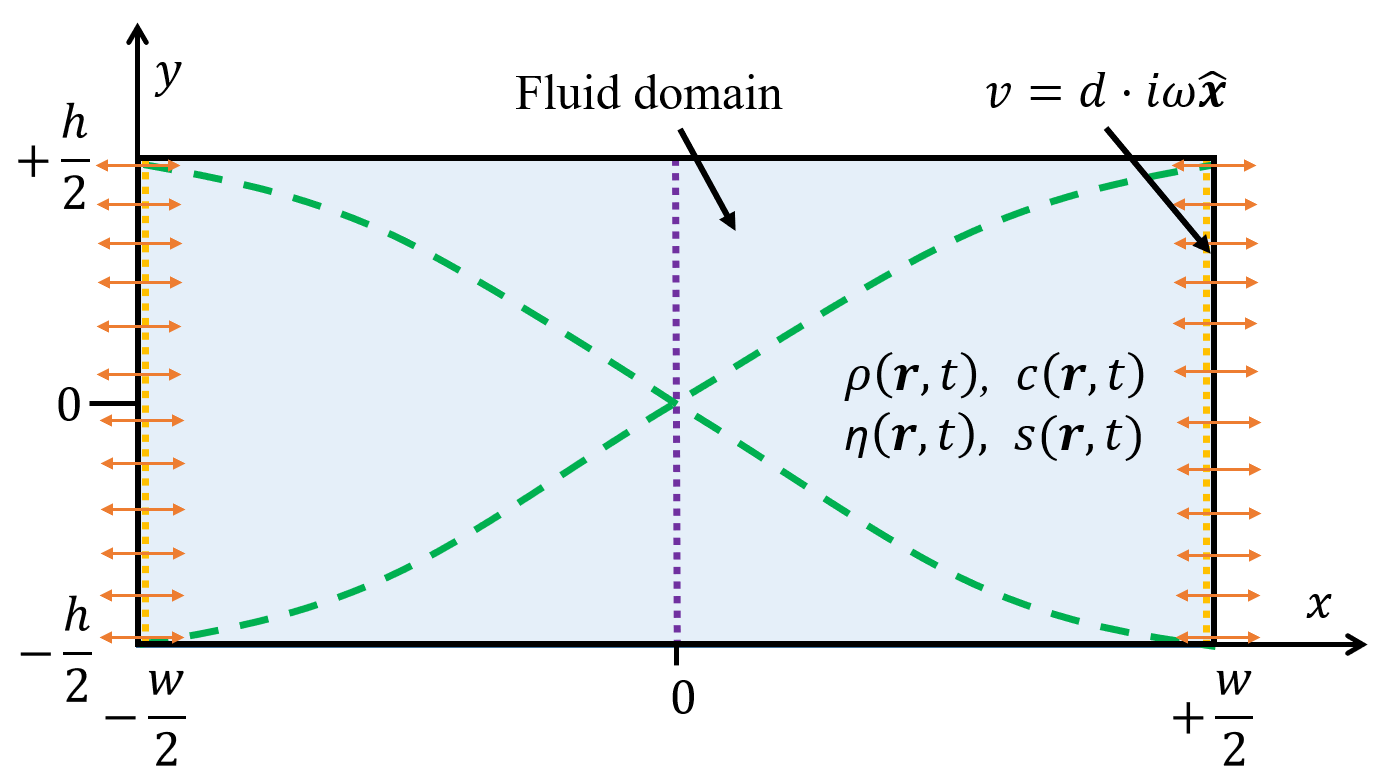} 
    \caption{Sketch of acoustofluidic microchannel with an imposed half-wave acoustic pressure resonance, containing fluids whose density, speed of sound, dynamic viscosity and solute concentration are functions of space and time.}
\label{Fig 1}
\end{figure}

 It is well-known that the above force is responsible for boundary-driven Rayleigh streaming and bulk-driven Eckart streaming in homogeneous fluids. In this letter, we proceed to show the same force is also responsible for recently observed streaming suppression and acoustic relocation of miscible as well as immiscible inhomogeneous fluids. A microchannel of width $w = 380$  $\mu m$ and height  $h = 160$ $\mu m$ containing an inhomogeneous miscible solution is chosen for study Fig.(\ref{Fig 1}). An acoustic standing half-wave along the width is imposed in the microchannel by actuating both the side walls at a specific frequency $'f'$ and wall displacement $'d'$ in x-direction. The first order fields due to this standing wave are obtained by solving the Eqs.(\ref{Eq 3}) in frequency domain, see Supplemental Material . These first order fields are then used to obtain the acoustic body force Eq.(\ref{Eq 7}) to solve the time averaged second order fields Eqs.(\ref{Eq 6}). At each time step, the first order fields are solved for the updated inhomogeneous fluid configuration ($\rho_0, c_0, \mu_0$) from the previous time step. Hence, both the first order and second order equations are bidirectionally coupled and solved numerically at all slow time-steps in COMSOL Multiphysics 5.6.

\begin{figure}[h!]
  \center
    \includegraphics[width=1\linewidth]{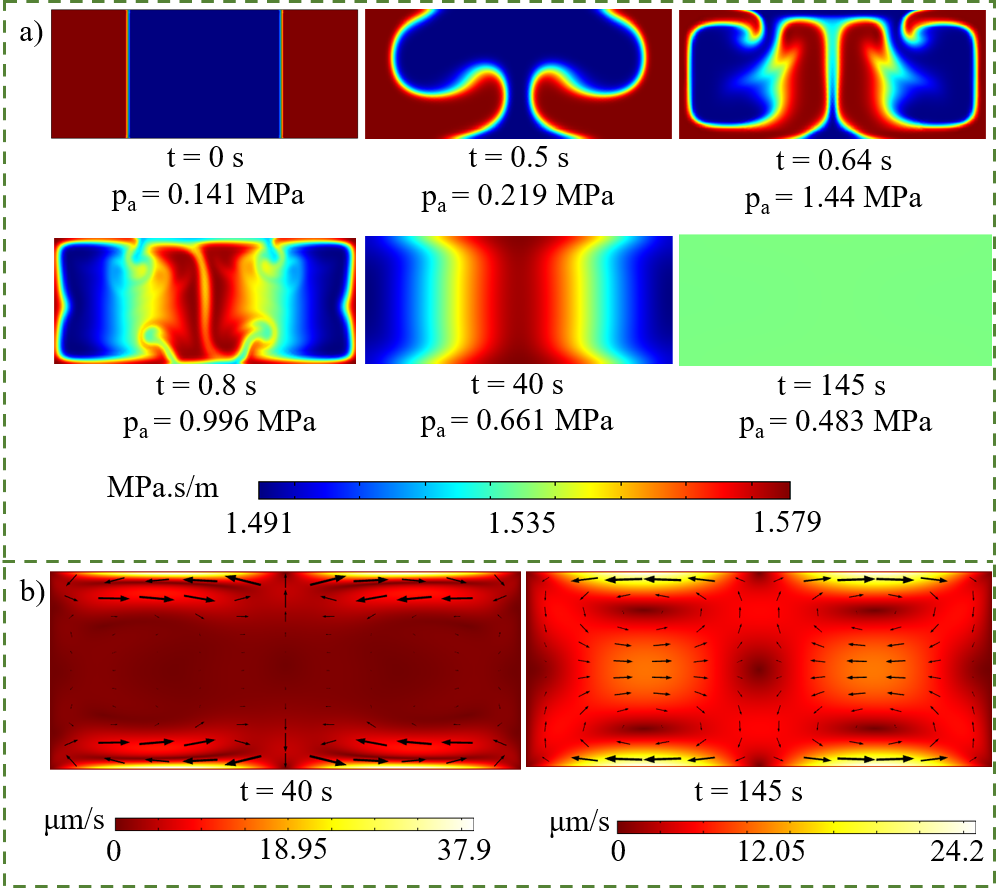} \caption{(a) Relocation of unstable configuration to stable configuration with $d=0.261$ nm and $f_0=1.96$ MHz. Initially (t=0s), low impedance DI Water (blue) at the center and high impedance 10\% Ficoll PM70 (red) at the sides. (b) second order velocity $\textbf{\emph{v}}_2$. Fluid properties are taken from \cite{Qiu2019Feb}}
\label{Fig 2}
\end{figure}

\begin{figure}[h!]
  \center
    \includegraphics[width=1\linewidth]{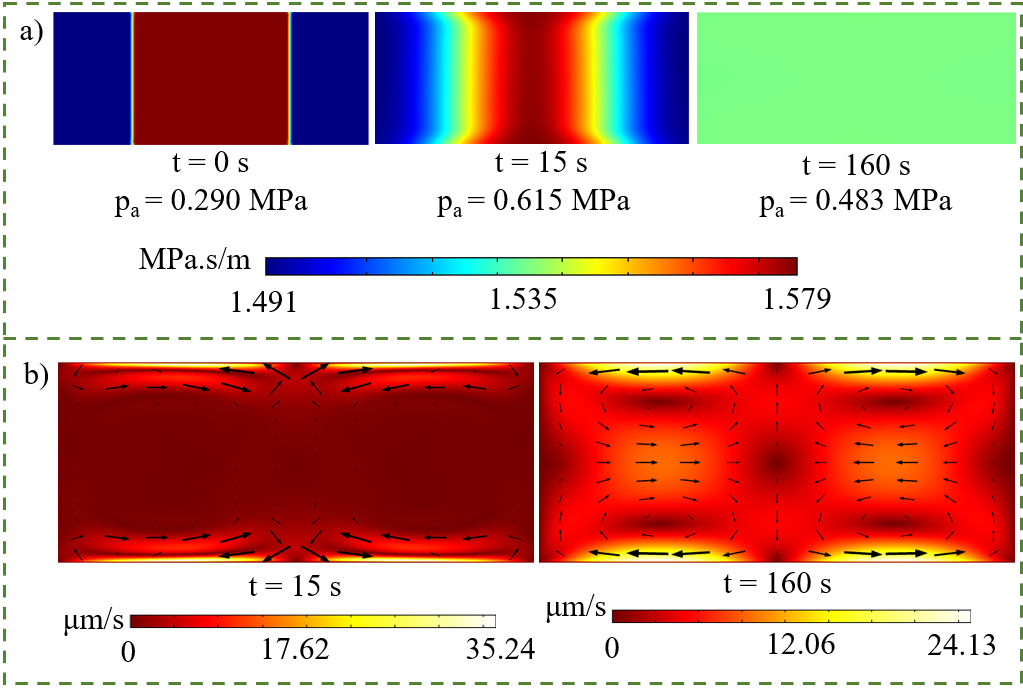} \caption{(a) Stable configuration with no relocation, low impedance DI Water (blue) at the sides and high impedance 10\% Ficoll PM70 (red) at the center (b) Second order velocity $\textbf{\emph{v}}_2$. Here, $d=0.261$ nm and $f_0=1.96$ MHz. Relocation force ($\textbf{\emph{f}}_{rl}$) stabilizes this configuration against gravity stratification.}
\label{Fig 3}
\end{figure}

\begin{figure}[h!]
  \center
    \includegraphics[width=1\linewidth]{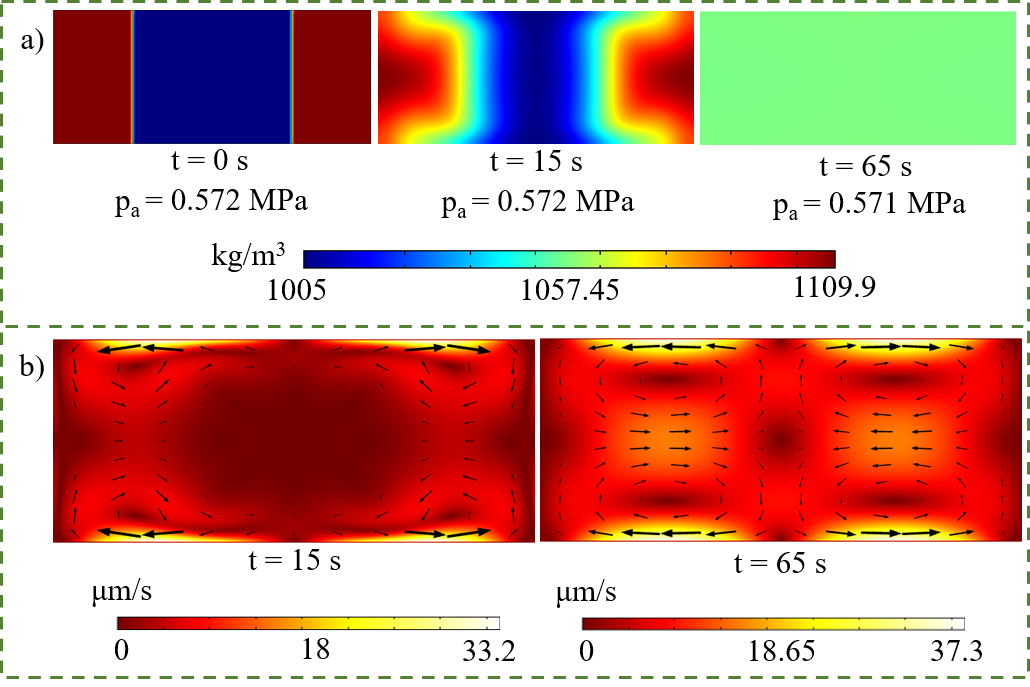} \caption{For constant impedance fluids any configuration of $\rho_0$ and $c_0$ is neutral configuration (a) No relocation due to zero relocation force ($\textbf{\emph{f}}_{rl} = 0$). To clearly show the absence of relocation force in constant impedance case, gravity is neglected. The presence of gravity would stratify the fluids. (b) Second order velocity $\textbf{\emph{v}}_2$. Here, $d=2$ nm and $f_0=1.96$ MHz.}
\label{Fig 4}
\end{figure}

\emph{Results.} - From Fig.(\ref{Fig 2}), acoustic body force $\textbf{\emph{f}}_{ac}$ tends to relocate high impedance fluid to node and low impedance to anti-nodes called stable configuration (whereas, any other configuration with impedance gradient is considered unstable). Once stable configuration is reached, due to existing impedance gradient, this $\textbf{\emph{f}}_{ac}$ inhibits any fluid motion due to gravity and suppresses acoustic streaming, that tries to disturb the stable configuration as seen in Figs.(\ref{Fig 2} \& \ref{Fig 3}). As time progresses, due to diffusion, the fluid profile becomes homogeneous, where the same $\textbf{\emph{f}}_{ac}$ induces boundary-driven Rayleigh streaming. From these results, it is evident that the acoustic body force $\textbf{\emph{f}}_{ac}$ is responsible for acoustic relocation and streaming suppression in inhomogeneous fluids as well as acoustic streaming in homogeneous fluids. Remarkably, in the process of acoustic relocation and diffusion as shown in Figs.(\ref{Fig 2}a and \ref{Fig 3}a), the amplitude of the first order fields ($p_a$ and $v_a$) vary significantly (refer Supplemental Material ) as the background $\rho_0$ and $c_0$ fields change in slow time-scale. Hence, it is now apparent that assumption of acoustic force density $E_{ac}$ to be constant is incorrect. Surprisingly, it is observed that for the case of constant impedance Fig.(\ref{Fig 4}), relocation does not occur irrespective of $\rho_0$ and $c_0$ configurations, thus any constant impedance configuration is called neutral configuration. This demonstrates that impedance gradient is the requisite and governing factor for acoustic relocation. Along with this, it is also observed from the simulations that, the sufficient conditions for acoustic relocation include - fluid interface should not be at node (Supplemental Material ) and the fluid configuration must be unstable.   
\par In order to explain the above results mathematically we analyse the acoustic body force, $\textbf{\emph{f}}_{ac}=-\nabla\cdot\langle\rho_0 \textbf{\emph{v}}_1 \textbf{\emph{v}}_1\rangle$ in detail
\begin{equation}
\label{Eq 8}
    \textbf{\emph{f}}_{ac}=-\nabla\cdot\langle\rho_0 \textbf{\emph{v}}_1 \textbf{\emph{v}}_1\rangle=-\rho_0\langle\textbf{\emph{v}}_1 \cdot\nabla\textbf{\emph{v}}_1\rangle-\langle\rho_1 \partial_t \textbf{\emph{v}}_1\rangle
\end{equation}

\par Using the following identity,  $\textbf{\emph{A}}\cdot\nabla\textbf{\emph{A}}=\nabla(\textbf{\emph{A}}^2/2)-\textbf{\emph{A}}\times(\nabla\times\textbf{\emph{A}})$ and substituting first-order fields from Eqs.(\ref{Eq 3}). Eq.(\ref{Eq 8}) becomes
\begin{multline}
\label{Eq 9}
   = -\frac{1}{2\rho_0}\nabla\langle|\rho_0 \textbf{\emph{v}}_1|^2\rangle+\langle\textbf{\emph{v}}_1\times\nabla\times(\rho_0 \textbf{\emph{v}}_1)\rangle+\frac{1}{2}\kappa_0\nabla\langle |p_1|^2\rangle\\-\langle(\kappa_0p_1)(\eta\nabla^2\textbf{\emph{v}}_1-\beta\eta\nabla(\nabla\cdot\textbf{\emph{v}}_1))\rangle
\end{multline}

\par For detailed derivation refer Supplemental Material . From the scaling analysis, last term in Eq.(\ref{Eq 9}) can be neglected, thus reducing to 
\begin{multline}
\label{Eq 10}
    =\frac{1}{2}\nabla(\kappa_0\langle 
    |p_1|^2\rangle-\rho_0 \langle|\textbf{\emph{v}}_1|^2\rangle)+\{\langle\textbf{\emph{v}}_1\times\nabla\times(\rho_0\textbf{\emph{v}}_1)\rangle\}
    \\
    -\frac{1}{2}[ \langle|p_1|^2\rangle\nabla\kappa_0+\langle|\textbf{\emph{v}}_1|^2\rangle\nabla\rho_0]
\end{multline}

\par In the above equation Eq.(\ref{Eq 10}), the first term consists of purely gradient term (conservative) and curl of that term is zero, thus does not cause relocation or streaming but only contributes to second order pressure. The third term within the square bracket is responsible for relocation, since the curl of this term in general is non-zero. From Eq.(\ref{Eq 3b}), the first order fields in the inviscid region satisfies the following relation, $\nabla\times(\rho_0 \textbf{\emph{v}}_1)=0$. Thus the term within the curly brackets is only significant inside the boundary layer ($\delta\sim 1\mu m$) of first order fields, which is responsible for boundary-driven streaming in inhomogeneous fluids. The competition between the relocation force (third term) and streaming force (second term) accounts for streaming suppression as seen in Figs.(\ref{Fig 2} and \ref{Fig 3}). In Eq.(\ref{Eq 10}), we have analytically separated relocation and streaming causing terms from the general body force term, $\textbf{\emph{f}}_{ac}$, which was previously claimed not possible \cite{Karlsen2018Jan}. For homogeneous fluids Eq.(\ref{Eq 10}) reduces to \cite{Friend2011Jun,Bradley1998Aug}
\begin{multline}
\label{Eq 11}
    \textbf{\emph{f}}_{ac}=-\nabla\cdot\langle\rho_0\textbf{\emph{v}}_1 \textbf{\emph{v}}_1\rangle =\frac{1}{2}\left(\kappa_0\nabla\langle 
    |p_1|^2\rangle-\rho_0\nabla\langle|\textbf{\emph{v}}_1|^2\rangle\right)
    \\
    +\{\rho_0\left\langle\textbf{\emph{v}}_1\times(\nabla\times\textbf{\emph{v}}_1)\right\rangle\}
\end{multline}
where the first term is homogeneous second order mean Eulerian pressure and the second term is responsible for acoustic streaming. 

\par To understand the second order pressure and relocation phenomenon in inhomogeneous fluids clearly, we study the acoustic body force outside the boundary layer of first order fields (neglecting streaming term). Thus, Eq.(\ref{Eq 10}) after time averaging reduces to 
\begin{multline}
\label{Eq 12}
    \textbf{\emph{f}}_{ac}=-\nabla\cdot\langle\rho_0\textbf{\emph{v}}_1 \textbf{\emph{v}}_1\rangle =\frac{1}{4}\nabla\left(\kappa_0 
    \langle|p_1|^2\rangle-\rho_0 \langle|\textbf{\emph{v}}_1|^2\rangle\right)
    \\
    \left[-\frac{1}{4}\langle|p_1|^2\rangle\nabla\kappa_0-\frac{1}{4}\langle|\textbf{\emph{v}}_1|^2 \rangle\nabla\rho_0\right] = \textbf{\emph{f}}_{1}+[\textbf{\emph{f}}_{2}]
\end{multline}
\par Considering the case of stable configuration Fig.(\ref{Fig 3}), second order velocity will be zero as relocation does not take place due to which Eq.(\ref{Eq 6b}) will reduce to $-\nabla\cdot\langle\rho_0\textbf{\emph{v}}_1 \textbf{\emph{v}}_1\rangle = \nabla\langle p_2\rangle$. Now, it is evident that in inviscid case, both the terms in Eq.(\ref{Eq 12}) contribute to the mean Eulerian pressure $\langle p_2\rangle$. Whereas, in \cite{Karlsen2016Sep,Karlsen2018,Karlsen2018Jan}, it is hypothesised that only first term in Eq.(\ref{Eq 12}) contributes to mean Eulerian pressure, $\langle p_2\rangle=\frac{1}{2}\kappa_0 \langle|p_1|^2\rangle-\frac{1}{2}\rho_0\langle|\textbf{\emph{v}}_1|^2\rangle$. Thus our work clearly demonstrates, the assumption on $p_{2}$ is needless and can be derived from the governing equations. 

\par In case of constant impedance ($Z_0=\rho_0 c_0 =constant$) inhomogeneous fluids, subjected to acoustic standing half wave ($\lambda \approx 2w$), using the relations $p_1=p_a sin(k{x})$ and $v_1=\frac{p_a}{i\rho_0 c}cos(k{x})$ where $k= 2\pi/\lambda$ is wave number. The relocation force term $\textbf{\emph{f}}_{2}$  reduces to 
\begin{multline}
\label{Eq 13}
    \textbf{\emph{f}}_{2|_{Z=C}}=-\frac{1}{4} |p_1|^2\nabla\kappa_0-\frac{1}{4}|\textbf{\emph{v}}_1|^2\nabla\rho_0 =  -\nabla\left(\frac{p_a^2\rho_0}{4Z^2}\right)
\end{multline}

\par Since $\textbf{\emph{f}}_{2}$ reduces to a purely gradient term in case of constant impedance inhomogeneous fluids, it does not induce relocation or motion but only contributes towards second order pressure. Thus, we prove that impedance gradient is the necessary condition for acoustic relocation, which agrees with experimental results \cite{Hemachandran2019Apr,Deshmukh2014Jul}.

\par For the case of variable impedance, the relocation term $f_2$ in Eq.(\ref{Eq 12}) is written as  
\begin{multline}
\label{Eq 14}
    \textbf{\emph{f}}_{2}=\frac{p_a^2 \sin^2{kx} }{4}\left(\frac{\nabla\rho_0}{\rho_0^2 c_0^2}\right) + \frac{p_a^2 \sin^2{kx}}{2}\left(\frac{\nabla c_0}{\rho_0 c_0^3}\right) - \\ \frac{p_a^2 \cos^2{kx}}{4}\left(\frac{\nabla\rho_0}{\rho_0^2 c_0^2}\right)
\end{multline}
\begin{equation}
\label{Eq 15}
    \textbf{\emph{f}}_{2}=-\frac{p_a^2\cos({2kx})}{4\rho_{0}^2 c_{0}^3}\nabla Z_0 + \frac{p_a^2}{4\rho_{0} c_{0}^3}\nabla c_0
\end{equation}

\par Analogous to Boussinesq approximation, $\frac{\nabla Z_0}{\rho_0^2 c_0^3}\approx\frac{\nabla Z_0}{\rho_{avg}^2 c_{avg}^3}, \frac{\nabla c_0}{\rho_0 c_0^3}\approx\frac{\nabla c_0}{\rho_{avg} c_{avg}^3}$ and substituting Eq.(\ref{Eq 15}) in Eq.(\ref{Eq 12}), separating gradient and non-gradient terms, 
\begin{multline}
\label{Eq 16}
    \textbf{\emph{f}}_{ac}=-\nabla\cdot\langle\rho_0\textbf{\emph{v}}_1 \textbf{\emph{v}}_1\rangle = -E_{ac}\cos({2kx})\nabla\hat Z_0 \\-\nabla\left(\frac{1}{4}\rho_0 |\textbf{\emph{v}}_1|^2-\frac{1}{4}\kappa_0
    |p_1|^2-E_{ac}\hat c_0\right) 
\end{multline}
where $E_{ac}=p_a^2/(4\rho_{avg} c_{avg}^2)$, $\hat c_0 = c_0/ c_{avg}$, $\hat \rho_0 = \rho_0/ \rho_{avg}$, and  $\hat Z_0 = \hat\rho_0\hat c_0$.

\par It is clear that only first term in Eq.(\ref{Eq 16}) is responsible for acoustic relocation in inhomogeneous fluids, whereas second term resembles conservative or purely gradient term, thus induces only pressure. Thus, in addition to impedance gradient being the necessary factor for relocation, now we write relocation force in terms of impedance gradient as follows
\begin{equation}
\label{Eq 17}
\textbf{\emph{f}}_{rl}=-E_{ac}\cos({2kx})\nabla\hat Z_0
\end{equation}


\par The above force term which is a part of generalised force equation Eq.(\ref{Eq 7}), is responsible for relocation of unstable configuration in Fig.(\ref{Fig 2}) and maintaining the stable configuration by inhibiting acoustic streaming as well as gravity stratification in Fig.(\ref{Fig 3}). Whereas, for constant impedance fluids (Fig.(\ref{Fig 4})), irrespective of any fluid configuration, relocation force is always absent ($\textbf{\emph{f}}_{rl} = 0$).

\par In several previous studies \cite{Qiu2019Feb,Karlsen2017Mar,Nath2019Nov}, including ours \cite{Pothuri2019Dec,Kumar2021Jul}, Eq.(\ref{Eq 15}) has been used to study relocation without realising that it contains implicit gradient or conservative terms which does not contribute to relocation. This leads to incorrect scaling analysis and thus it is necessary to ignore this gradient term in such circumstances. Furthermore, Eq.(\ref{Eq 17}) captures all the aspects of relocation and does not cause scaling issues like Eq.(\ref{Eq 15}). It must be noted that Eq.(\ref{Eq 17}) is valid only outside the boundary layer of first order fields, whereas inside the boundary layer, it cannot be reduced from $\textbf{\emph{f}}_{2}$ of Eq.(\ref{Eq 12}). This is because $p_1=p_a sin(kx)$ remains same inside and outside the boundary layer ($\delta\sim 1\mu m$), but $v_1\neq\frac{p_a}{i\rho_0 c}cos(kx)$ and $\partial_y v_1$ is significant inside the boundary layer due to no slip condition. Thus, in case of constant impedance fluids, relocation force is absent in the bulk but non-zero within the boundary layer, which is responsible for the disturbance of homogeneous streaming as in Fig.(\ref{Fig 4}).

\begin{figure}[H]
  \center
    \includegraphics[width=1\linewidth]{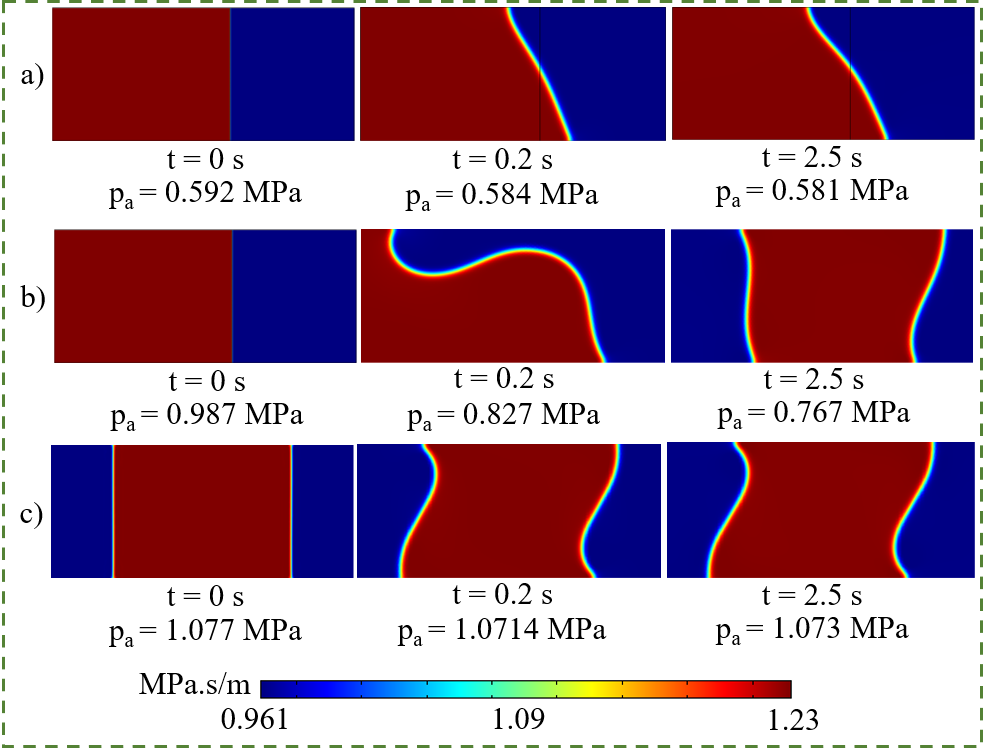} \caption{Relocation of immiscible fluids consisting high impedance mineral oil(red) and low impedance silicone oil(blue) with a surface tension of $1$ $mN/m$ actuated at a frequency of $2.1$ $MHz$ for unstable configurations a) No relocation for $E_{ac} < E_{cr}$, b) Relocation for $E_{ac} > E_{cr}$ and c) stable configuration - No relocation for any $E_{ac}$}     \label{Fig 5}
\end{figure}

\emph{Immiscible fluids}.- The phenomenon of acoustic relocation of immisicble fluids due to standing acoustic wave is also governed by Eq.{\eqref{Eq 7}}. This theory predicts that for the relocation of immiscible fluids to occur, the applied acoustic energy density $E_{ac}$ must be greater than the threshold $E_{cr}$ in order to overcome the interfacial tension force. The above prediction is in agreement with the experimental studies by Hemachandran et al. \cite{Hemachandran2019Apr}. However, in their study relocation is achieved irrespective of the interface location with respect to node. Also, the frequency employed is much different than the resonant frequency ($f=c_{avg}/2w$) that corresponds to the standing half-wave actuated only along the width of microchannel as shown in Fig.(\ref{Fig 1}). On contrary to their study, when we actuated only side walls, relocation seemed to be highly dependent on the position of the interface and was absent when the interface and node coincide. Whereas, when all the walls were actuated in direction to their surface normal, we achieved relocation, independent of the interface location, (Fig.(\ref{Fig 5})) that is in agreement with their experiments \cite{Hemachandran2019Apr}. It is evident that relocation can be achieved through two modes; one is due to 1D standing wave resulting from the actuation of side walls at 1D resonant frequency ($f=c_{avg}/2w$) while the other is due to 2D standing wave resulting due to actuation of all walls at 2D resonant frequency (between $f=c_{avg}/2w$ and $f=c_{avg}/2h$). Since the relocation in \cite{Hemachandran2019Apr} is due to 2D standing wave, the frequency required ($\sim 2.1-2.4 MHz$) for relocation (for silicone-mineral oil) is different from 1D resonant frequency ($1.66 MHz$). 
\\
\par\emph{Conclusion} - We have put forward a theory of nonlinear acoustics that governs the acoustic phenomena of relocation and streaming suppression in inhomogeneous miscible and immiscible fluids (including streaming in homogeneous fluids). This theory also confirms the fact that divergence of time-averaged Reynolds tensor $-\nabla\cdot\langle\rho_0\textbf{\emph{v}}_1\textbf{\emph{v}}_1\rangle$ is alone responsible for all the above processes. We showed that first order fields $p_1, \textbf{\emph{v}}_1$ and energy density $E_{ac}$ vary significantly during process of acoustic relocation and diffusion. Importantly, we have proved that impedance gradient is the necessary condition for relocation. The other  conditions for acoustic relocation in 1D and 2D mode will be addressed in detail using stability analysis for both miscible and immiscible fluids in upcoming paper. The fundamental understanding from this study can give new insights for particle (cells/drops/beads)  and inhomogeneous fluid handling in microchannel under acoustic fields.

\nocite{*}
\bibliography{references}
\end{document}